\documentclass[letterpaper, 10 pt, conference]{ieeeconf}  % Comment this line out if you need a4paper

\IEEEoverridecommandlockouts                              % This command is only needed if 
\usepackage{amsmath}
\usepackage{graphicx}
\usepackage{float}
\usepackage{listings}
\usepackage{algorithm}
\usepackage{algpseudocode}
\usepackage{amsfonts}
\usepackage{xcolor}
\usepackage{cite}

\newtheorem{remark}{Remark}
\usepackage{subcaption}

\title{\LARGE \bf
Data-Driven Deep Learning Based  Feedback Linearization \\ of Systems with Unknown Dynamics
}

\author{Raktim Gautam Goswami$^{1}$, Prashanth Krishnamurthy$^{1}$, Farshad Khorrami$^{1}$% <-this % stops a space
\thanks{$^{1}$Control/Robotics Research Laboratory, Elec. \& Comp. Engg. Dept., Tandon School of Engineering (Polytechnic Institute), New York University, Brooklyn, NY, 11201
        {\tt\small \{rgg9769,pk929,khorrami\}@nyu.edu} \newline This work was supported in part by ARO grant  W911NF-22-1-0028 and in part by the New York University Abu Dhabi (NYUAD) Center for Artificial Intelligence and Robotics, funded by Tamkeen under the NYUAD Research Institute Award CG010.}
}

\begin{document}

\maketitle
\thispagestyle{empty}
\pagestyle{empty}

%%%%%%%%%%%%%%%%%%%%%%%%%%%%%%%%%%%%%%%%%%%%%%%%%%%%%%%%%%%%%%%%%%%%%%%%%%%%%%%%
\begin{abstract}

A methodology is developed to learn a feedback linearization (i.e., nonlinear change of coordinates and input transformation) using a data-driven approach for a single input control-affine nonlinear system with unknown dynamics. 
% The approach is also applicable to uncertain nonlinear systems when a feedback linearizing controller is available based on the nominal system dynamics; however, parametric and dynamic uncertainties will result in residual nonlinear dynamics that may be learned in real-time.
We employ deep neural networks to learn the feedback law (input transformation) in conjunction with an extension of invertible neural networks to learn the nonlinear change of coordinates (state transformation). We also learn the matrices $A$ and $B$ of the transformed linear system and define loss terms to ensure controllability of the pair $(A,B)$. The efficacy of our approach is demonstrated by simulations on several nonlinear systems. Furthermore, we show that state feedback controllers designed using the feedback linearized system yield expected closed-loop behavior when applied to the original nonlinear system, further demonstrating validity of the learned feedback linearization.

\end{abstract}

%%%%%%%%%%%%%%%%%%%%%%%%%%%%%%%%%%%%%%%%%%%%%%%%%%%%%%%%%%%%%%%%%%%%%%%%%%%%%%%%
\section{INTRODUCTION}
To benefit from the wealth of analysis and design tools for linear systems, feedback linearization of nonlinear systems has received significant attention over many decades \cite{krener,isidori}.
Feedback linearization is a geometric approach  to obtain a linear dynamical representation of a system by means of a nonlinear transformation of the state (i.e., a change of coordinates) and input (i.e., a nonlinear feedback). While feedback linearization has been widely applied \cite{fl-ap1,fl-ap2,fl-ap3}, the requirement of knowledge of the system dynamics can be a constraint in practice since many real-world phenomena (e.g., friction, backlash) are difficult to accurately model and real-world systems are subject to parametric uncertainties, necessitating adaptive approaches \cite{adap-fl1,adap-fl2,adap-fl3, extended_observer_fl}. In this work, we develop a data-driven approach using deep learning to learn the feedback law and nonlinear change of coordinates to feedback linearize single input control-affine nonlinear systems with unknown dynamics.

In recent years, machine learning (and, in particular, deep learning) has emerged as a powerful tool in many applications. The availability of several open-source packages has also facilitated the implementation of deep learning systems. 
% Especially, deep learning has gained popularity because of its approximation capability of complex functions. 
% Also, several open source packages have been developed to increase the ease of implementation of deep learning algorithms.  
In control systems, deep learning and other data-driven methods are often collectively referred to as data-driven control and are widely used in applications such as system identification, trajectory tracking, and optimal controller design \cite{ddc-survey,bolun-paper,dl-system-id,ddc-survey2}.

Neural network approaches to learn feedback linearizing controllers have been studied assuming the system is in the normal form (i.e., chain of integrators with the last state containing the nonlinearities). In such cases, there is no need for a nonlinear change of coordinates. In
 \cite{fl-nn}, neural networks are used to linearize systems in normal form.  
 %the  to track an output using a feedback linearization based controller for a system in controllable canonical form with unknown input-affine nonlinearities at the gradient of the last state of the system. 
  \cite{fl-gaussian-processes} used Gaussian processes (a nonparametric Bayesian approach) to identify the nonlinear functions in dynamics of the last state of the normal form and use the identified functions to construct a feedback linearizing control law to make the system globally uniformly ultimately bounded.  Improving on this, \cite{fl-online-gp} addressed simultaneous identification and control based on updating the controller using event triggers when the uncertainty becomes large. %However, \cite{fl-nn}, \cite{fl-gaussian-processes}, and \cite{fl-online-gp} do not address the problem of feedback linearizing a system which is not in the controllable canonical form.
  
In \cite{lfl}, a discrete-time system was considered and a method was developed to learn a mapping to predict the control input at a sampling time from the current and target next state vector and it was shown that this learned mapping can be used to implement a proportional-integral (PI) controller. A model-free reinforcement learning methodology was developed in \cite{berkeley-paper} to construct an output tracking linearizing controller for a system with uncertain dynamics. Considering a nonlinear non-input-affine discrete-time system, \cite{fl-nonaffine} proposed an algorithm to convert the system to an approximate input-affine form and used a feedback linearization based controller to track an output signal. However, the learning of a state transformation for obtaining a feedback linearized system representation  was not considered in \cite{fl-nonaffine,berkeley-paper}. Apart from the controls literature, methods have been developed to approximate nonlinear ordinary differential equations (ODEs) with linear ones using methods such as Koopman operator along with auto-encoders \cite{nature-paper} and invertible neural networks \cite{INN} with a base ODE \cite{learning-ode}.

We propose a deep learning based methodology (Section \ref{prop-sol}) for input-state feedback linearization of an uncertain nonlinear system.
We learn a nonlinear change of coordinates (using an extension of invertible neural networks \cite{INN}) and input transformation (using feedforward neural networks) to transform the uncertain system into a linear dynamic model (with $A$ and $B$ matrices also being learned). 
% We use feedforward neural networks and an extension of invertible neural networks \cite{INN} respectively for finding these functions. 
%% We show that suitable matrices for the feedback linearized system can be found in the same training approach.
We define loss function terms to model the accuracy of the learned feedback linearization, controllability of the resulting linear system, and boundedness of the new input.
% For ensuring controllability of this linear system, we define an additional loss function term using the controllability matrix. 
We show the efficacy of our approach in simulation studies (Section \ref{simul}) through evaluations of errors between trajectories of the original system and predicted trajectories obtained from the learned feedback linearized representation. 
% The rest of the paper is structured as follows. 
%The considered problem of feedback linearization for systems with unknown dynamics is formulated in Section \ref{prob-form}. 
%% Our proposed algorithmic solution is presented in Section \ref{prop-sol}. In Section \ref{simul}, we validate our approach through simulations on several nonlinear systems. 
% We end by providing an analysis on the stability of the feedback linearized system and show that desired closed loop behaviour of the original nonlinear system can be achieved using the learned controller.

\section{PROBLEM FORMULATION}
\label{prob-form}
\label{section:prob}
\subsection{Input to State Feedback Linearization}
\label{sec:i2sfl}
Consider a nonlinear system in the input-affine form, $\dot{x} = f(x) + g(x)u$,
% \begin{equation}
%     \dot{x} = f(x) + g(x)u
%     \label{nonlinear-system}
% \end{equation}
with state $x \in \mathbb{R}^{n}$, input $u \in \mathbb{R}$, $f:\mathbb{R}^{n} \to \mathbb{R}^{n}$, and $g:\mathbb{R}^{n} \to \mathbb{R}^{n}$.  Input to state feedback linearization involves finding a nonlinear control law and a change of coordinates in the form
\begin{align}
u &= \alpha(x) + \beta(x)v \ \ ; \ \ 
z =\phi(x)
\label{fl1_fl2}
\end{align}
so that in the new coordinates $z$ (with $v \in \mathbb{R}$ being the new control input), the system dynamics have a linear structure (e.g., Brunovsky's form) where $(A,B)$ is a controllable pair:
\begin{equation}
     \dot{z} = Az + Bv.
     \label{linear-system}
\end{equation}
One approach for constructing an input to state feedback linearization \cite{isidori} is to find an output map $\hat{y} = {h}(x)$, ${h}:\mathbb{R}^{n} \to \mathbb{R}$ such that the relative degree of $\hat{y}$ with respect to input $u$ is $n$. 
% Then, with the standard Lie algebra notation (e.g., $L_f h(x)=\frac{\partial h}{\partial x}f(x)$), we have
% \begin{align}
% %\begin{split}
%     \dot{\hat{y}} = \frac{\partial h(x)}{\partial x}(f(x) + g(x)u)
%                 = L_f{h}(x) + L_g{h}(x)u.
% %\end{split}
% \end{align}
% % where $L_f{{h}}(x)$ denotes the Lie derivative of $ h(x)$ along the vector field $f$. 
% Since the relative degree of system with output $\hat{y}$ and input $u$ is $n$, we have $L_g{h}(x)u = 0$ if $n>1$. Similarly, for all $\gamma <n$, $L_gL_f^{\gamma-1}{h}(x)u = 0$. We can thus show that
With such an output, we have
\begin{equation}
    \hat{y}^{(\gamma)} = 
   \left\{
        \begin{array}{ll}
            L_f^{\gamma}{h}(x) & \quad \gamma < n \\
            L_f^{n}{h}(x) + L_gL_f^{n-1}{h}(x)u & \quad \gamma = n,
        \end{array}
    \right.
\end{equation}
where $L_f$ and $L_g$ represent the Lie derivatives with respect to $f$ and $g$, respectively.
 Applying $u = \frac{v - L_f^{n} h(x)}{L_gL_f^{n-1}{h}(x)}$, we obtain $\hat{y}^{(n)} = v$. Setting $z_1 = \hat{y}$, $z_2 = \hat{y}^{(1)}$, \dots, $z_{n} = \hat{y}^{(n-1)}$, the dynamics of the system in the new coordinates $z=[z_1, \dots, z_n]^T$ are of the form  (\ref{linear-system}).

\subsection{Systems with Model Uncertainties}
\label{subsec:model-uncert}
In practice, the system model might be unknown or only approximately known. While an approximate feedback linearization might be achievable given a nominal model of a system, perfect feedback linearization might be unattainable under model uncertainties. When the system model is completely unknown, even an approximate feedback linearization is not analytically possible. In such cases, our proposed data-driven approach can be applied to use empirical (input, state) data from the uncertain system to {\em learn} a feedback linearization. 
% We consider a general scenario where the system dynamics are completely unknown with the only available information being the system order and with the system state considered to be measurable.
We consider a general scenario where the system dynamics are completely unknown with the only available information being the system order. 
Also, by problem formulation for input-state feedback linearization, the system states are assumed to be measurable. 
% This is a rather mild assumption because in many physical systems, we can measure the states using sensors without knowing the system dynamics, e.g., angles and angular velocities of a robotic arm.
% Our proposed approach is based on the application of deep neural networks to learn a nonlinear control law and nonlinear change of coordinates using a data-driven approach to achieve a feedback linearization of the system.
% Since we do not assume any knowledge of the system dynamics, our proposed solution is robust to system modeling uncertainties.
Note that during the training phase, one can apply small inputs for shorter durations of time so that the system trajectories remain bounded (for an unknown system particularly if it is unstable). Additionally, one may apply a pre-feedback to stabilize the system (as commonly done in system identification problems  \cite{closed-loop-id1}, \cite{closed-loop-id2}).

%To achieve this, we need to simultaneously find
%\begin{itemize}
%    \item a control input
%    \begin{equation}
%   u = \alpha(x) + \beta(x)v
%    \end{equation}
%     which can feedback linearize the nonlinear system in (\ref{nonlinear-system})
%    \item a bijective nonlinear transformation $z = \phi(x)$ which can map the states $x$ of the nonlinear system to the states of a linear system $\dot{z} = Az + Bv$.
%    \item a controllable $(A,B)$ pair for the above linear system
%\end{itemize}

\subsection{Non-uniqueness of Feedback Linearization}
With the control law and change of coordinates given in (\ref{fl1_fl2}), the original system with state $x$ can be written in terms of the transformed system with state $z$ where $\phi(x) = z$. With $\phi$ being an invertible function (as required for any valid change of coordinates), we can write  $\phi^{-1}(z) = x$. Therefore,
    $\frac{\partial \phi^{-1}(z)}{\partial z}\dot{z} = \dot{x}$
from which we can write
\begin{equation}
\label{non-unique}
    \frac{\partial \phi^{-1}(z)}{\partial z}\left (A z + B \left (\frac{u-\alpha(x)}{\beta(x)}\right ) \right) = f(x) + g(x)u. \\
\end{equation}

The right-hand side of the above equation involves only functions from the original system. However, the left-hand side of the equation is a combination of the functions $\phi, \alpha, \beta$, and the matrices $A$ and $B$ which are all functions/matrices to be determined for the purpose of achieving the desired feedback linearization. It is easy to see that there can be more than one solution of these functions/matrices that satisfy (\ref{non-unique}). For instance, if $(\phi^*,\alpha^*,\beta^*,A^*,B^*)$ is a solution to (\ref{non-unique}), $(\phi^*,\alpha^*+K\phi^*\beta^*,\mu\beta^*,A^*+B^*K,\mu B^*)$ is also a valid solution for some $n$-dimensional row vector $K$, $\mu \in \mathbb{R}$, and $\mu \neq 0$.
This non-uniqueness makes it impossible to directly compare the learned functions/matrices with any analytical solution. Therefore, we use comparisons of the system trajectories as discussed in the subsequent sections as a metric for the accuracy of the learned solution.

\section{PROPOSED APPROACH}
\label{prop-sol}
\subsection{Algorithm}
Accomplishing feedback linearization can be viewed as finding the functions $\alpha, \beta, \phi$, and the matrices $A$ and $B$. We approximate $\alpha$ and $\beta$ by learnable neural networks $\alpha_{\theta}$ and $\beta_{\theta}$  each of which consists of sequences of fully-connected layers with Sigmoid Linear Unit activation ($silu$)  \cite{silu} between the layers:
    $silu(a) = \frac{a}{1 + e^{-a}}$.
 As $\beta_{\theta}$ appears in the denominator in $v = \frac{u-\alpha_{\theta}}{\beta_{\theta}}$, it is important that $\beta_{\theta} \neq 0$ to obtain a valid input transformation in the feedback linearization. This is realised by applying a nonlinear function $p(a) = sgn(a) \max(|a|,\epsilon_1)$ on the output layer of $\beta_\theta$ 
%  \begin{align}
%      p(a) = \left\{
%         \begin{array}{ll}
%             a & \quad |a| > \epsilon_1 \\
%             \epsilon_1 & \quad |a| \leq \epsilon_1,
%         \end{array}
%     \right.
%  \end{align}
 % \begin{align}
 %     p(a) = sgn(a) \max(|a|,\epsilon_1)
 % \end{align}
with $0<\epsilon_1<<1$. This does not entail any loss of generality since $\beta_\theta$ can be freely scaled (scaling of $\beta_\theta$ simply results in scaling of $v$ by the inverse of the scaling factor).

The basic essence of the proposed algorithm is as follows. We apply randomly generated inputs $u$ to collect trajectories of the system states $x$, with which we estimate the feedback linearized system trajectories $z$ using the above neural networks with parameters $\theta$. With these estimated $z$, we use $\phi_\theta^{-1}$ to obtain the estimated $x$ trajectories and use the difference between the original and estimated $x$ trajectories as part of the loss function to update $\theta$. Details of the utilized loss function are in Section \ref{subsec:learning}. This methodology is applied iteratively until the convergence of the loss function.

We estimate $\phi$ by $\phi_{\theta}$ which uses the invertible architecture of Section \ref{subsec:INN}. Learnable parameters $(A_{\theta}, B_{\theta})$ are used to estimate the discrete-time matrices ($A_d,B_d$) arising from discretization of the continuous-time system \eqref{linear-system} as 
% $z((k+1)T) = A_d \; z(kT) + B_d \; v(kT)$
\begin{align}
    z((k+1)T) = A_d \; z(kT) + B_d \; v(kT)
\end{align}
with $A_d = e^{AT}$ and $B_d = \int_{0}^{T} e^{A(T - \tau)} B d\tau$.  Since we need to collect trajectories and evaluate mismatches between observed and predicted trajectories, it is beneficial to work with a discretization for lower computational complexity.
While $\alpha_\theta$, $\beta_\theta$, and $\phi_\theta$ are functions of the state $x$, $A_\theta$ and $B_\theta$ are constant matrices to be learned during training.
For notational convenience, we denote the weights of all networks as a single vector $\theta$.

 %The continuous-time system $\dot{z} = Az + Bv$ may be replaced with its discretized version as 

% It is computationally more efficient to use the discretized dynamics for training.
% The learned dynamics, namely $(A_\theta, B_\theta)$, can be more efficiently computed during training using the discretized dynamics. 
%Moreover, sampling trajectories of $x$ and evaluating the loss functions are easier when working with the discretized form. Because of this, we find it useful in using the pair $(A_\theta, B_\theta)$ to estimate coefficients of the discretized systems, $(A_d,B_d)$, instead of the continuous system coefficients $(A,B)$.

\subsection{Learning $\alpha_\theta$, $\beta_\theta$, $\phi_\theta$, $A_\theta$, $B_\theta$}
\label{subsec:learning}
\renewcommand{\algorithmicrequire}{\textbf{Input:}}
\renewcommand{\algorithmicensure}{\textbf{Output:}}
\begin{algorithm}[!b]
\caption{Learning Feedback Linearization}
\label{alg:flus}
\begin{algorithmic}[1]
\Require $u = \{u(t_{j}), \dots, u(t_{j+m}) \}$, $x = \{x(t_{j}), \dots, x(t_{j+m}) \}$, $\theta$
% \Require  $\mathbb{D}_q$, $\theta$ 
\Ensure Updated $\theta$, $x_\theta=\{x_\theta(t_{j+1}),\dots, x_\theta(t_{j+m})\}$
% \State $x_\theta(t_0) \leftarrow x(t_0)$
\State \textbf{Prediction:}
\State $z(t_j) = \phi_\theta(x(t_j))$
\For{$i = j$ to $j+m-1$}
\State $v(t_i) = \frac{u(t_i)-\alpha_\theta(x(t_i))}{\beta_\theta(x(t_i))}$
\State $z(t_{i+1}) = A_\theta z(t_i) + B_\theta v(t_i) $
\State $x_\theta(t_{i+1}) = \phi_\theta^{-1}(z(t_{i+1}))$
\EndFor
\State \textbf{Update:}
\State Update $\theta$ using gradient descent on $\mathcal{L}$ in Eqn. (\ref{loss})
% \State Calculate loss $\mathcal{L}$ \Comment{loss function in (\ref{loss})}
% \State Calculate gradients of $\mathcal{L}$ w.r.t. (with respect to) $\theta$ 
% \State Update $\theta$ using gradient descent
\end{algorithmic}
\end{algorithm}

% Given an input $u(t)$, $t = (t_0,\dots,t_k)$, and an initial state $x(t_0)$ of the nonlinear system, we can estimate the states $x(t)$ using the method outlined in algorithm \ref{alg:flus}.

\begin{figure}
    \centering
    \includegraphics[scale = 0.4]{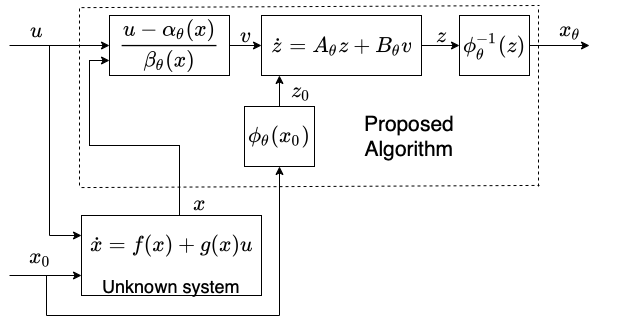}
    \caption{Block diagram showing the prediction step in Alg. \ref{alg:flus}.}
    \label{fig:block}
\end{figure}

With inputs $u$ (generated using the structure shown in Section \ref{subsec:inpsig}), we sample many trajectories of the states and store them in the data set $\mathbb{D}$ given by
\begin{equation}
\label{eqn:dataset}
    \mathbb{D} = \{(u_\tau,x_\tau)| \tau = \{t_0, t_1, \dots\},t_{k} = t_0 + kT\},
\end{equation}
with $T$ being the sampling time and $u_\tau$ and $x_\tau$ denoting samples of $u$ and $x$ at time instant $\tau$. For each iteration of training, we divide $\mathbb{D}$ into subsets $\mathbb{D}_q$, ($q=\{1,2,\dots\}$) with each $\mathbb{D}_q$ being of form $\mathbb{D}_q = \{({u_\tau},{x_\tau})|\tau = \{t_j,\dots, t_{j+m}\},t_{j+k} = t_j + kT\}$ with the value of $j$ (for each $\mathbb{D}_q$) being chosen as a non-negative integer less than the cardinality of $\mathbb{D}$. We initialize $\theta$ as uniform random values for the neural networks and Brunovsky's form for matrices $A_\theta$ and $B_\theta$. With each $\mathbb{D}_q$ as input, we update $\theta$ using Algorithm \ref{alg:flus}. To generate data \eqref{eqn:dataset}, we consider randomly sampled initial conditions (typically drawn to span ranges of particular relevance for the specific system). The overall flow of the prediction step of the Alg. \ref{alg:flus} is shown in Fig. \ref{fig:block}.

We want $x_{\theta}$ calculated in Alg. \ref{alg:flus} to approximate $x$. To this end, we define the objective function as 
\begin{equation}
\label{l1}
    \mathcal{L}_1 = \frac{1}{m}\sum_{i=j+1}^{j+m} \|x_\theta(t_i) - {x_{t_i}} \|_2^2.
\end{equation}
To ensure controllability of the learned pair $(A_\theta, B_\theta)$, the controllability matrix $\Gamma = [B_\theta, A_\theta B_\theta,\dots, A_\theta^{n-1} B_\theta]_{n\times n}$
needs to have full rank, which is satisfied by ensuring that $\det\Gamma$ is non-zero. Hence, we define a loss function as \begin{equation}
\label{l2}
    \mathcal{L}_2 = \frac{1}{\min(|\det \Gamma|,\epsilon_2)}
\end{equation}
with $0<\epsilon_2<<1$.
\begin{remark}
\label{det_lemma}
    Given any controllable pair ($A_\theta,B_\theta$), $\Gamma$ has full rank; therefore, $|\det \Gamma| \geq \epsilon_2$, if $\epsilon_2$ small enough. Therefore, $\mathcal{L}_2 = \frac{1}{\epsilon_2}$ implying the gradient of $\mathcal{L}_2$ w.r.t. $\theta$ is 0 for controllable pair ($A_\theta,B_\theta$).
\end{remark}
% \begin{proof}
%     For controllable pair $(A_\theta, B_\theta)$, $\Gamma$ has full rank; therefore, 
%     $|\det \Gamma| \geq \epsilon_2$, if $\epsilon_2$ is small enough. Thus, $\mathcal{L}_2 = \frac{1}{\epsilon_2}$ implying gradient of $\mathcal{L}_2$ w.r.t. $\theta$ is 0. 
% \end{proof}

As the ranges of $\alpha_\theta$ and $\beta_\theta$ have no constraints, the value of $v = \frac{u-\alpha_\theta(x)}{\beta_\theta(x)}$ can effectively become very large. In such cases, the dependence of $v$ on $u$ can become small depending on the range of values taken by $u$. Moreover, large values of $v$ will make the states $z$ of the linear system become very large. These behaviors are not desired. Therefore, to check this, we add a third loss function component as
\begin{equation}
\label{l3}
    \mathcal{L}_3 = \frac{1}{m}\sum_{i=j+1}^{j+m}\| \max(|v(t_i)|,\epsilon_3) \|
\end{equation}
where the constant $\epsilon_3>0$ is chosen empirically to be larger than (but comparable to) the maximum absolute value of the signal $u$ used in training and considered representative of input ranges in closed-loop operation in the actual system.
\begin{remark}
\label{v_lemma}
    Given $\epsilon_3 > 0$ and $-\epsilon_3 \leq v(t_i) \leq \epsilon_3$ $\forall i = j+1,\dots,j+m$, $\max(|v(t_i)|,\epsilon_3) = \epsilon_3 \implies \mathcal{L}_3 = \epsilon_3$. Therefore, the gradient of $\mathcal{L}_3$ w.r.t. $\theta$ is 0 when $-\epsilon_3 \leq v(t_i) \leq \epsilon_3$. 
\end{remark}

% \begin{proof}
%     For $-\epsilon_3 \leq v(t_j) \leq \epsilon_3$, $\max(|v(t_i)|,\epsilon_3) = \epsilon_3 \implies \mathcal{L}_3 = \epsilon_3$. Therefore, its gradient w.r.t. $\theta$ is 0.
% \end{proof}

Remarks \ref{det_lemma} and \ref{v_lemma} are important as they ensure that weights of the neural networks are affected by $\mathcal{L}_2$ and $\mathcal{L}_3$ only when $\Gamma$ and $v(t_i)$ are not in the desired ranges.
A linear combination of loss functions (\ref{l1})-(\ref{l3}) is defined as
\begin{align}
\label{loss}
    \mathcal{L} = a_1\mathcal{L}_1 + a_2\mathcal{L}_2 + a_3\mathcal{L}_3
\end{align}
with $a_1,a_2,a_3$ being real positive constants. These coefficients are selected so that $a_2,a_3<<a_1$ to weight $\mathcal{L}_1$ higher since $\mathcal{L}_1$ models the primary objective of the learning. 

\begin{remark}
As an alternate training strategy, we can divide the learning procedure above into two parts. In the first part, we learn the linearizing controller for a zero-input system as 
$v = \frac{-\alpha_\theta(x)}{\beta_\theta(x)} = \gamma_\theta(x)$.
In the second part, we fix $\gamma_\theta$ and learn $\beta_\theta(x)$ using the dataset $\mathbb{D}$ with
$v = \frac{u}{\beta_\theta(x)}+\gamma_\theta(x)$.
\end{remark}

\subsection{Invertible Neural Networks and Extensions}
\label{subsec:INN}
For creating an invertible architecture for estimating the bijective nonlinear mapping $\phi(.)$, we use an extension of the Invertible Neural Network (INN) architecture proposed by \cite{INN}. We split the $n$-dimensional input vector $x$ into $n$ components as $x_1,x_2, \dots, x_n$ and calculate the output as
{
\begin{align}
\begin{split}
    &z_1 = x_1  \exp(s_1(x_2,\dots,x_n)) + t_1(x_2,\dots,x_n) \\
    &z_2 = x_2  \exp(s_2(x_3,\dots,z_1)) + t_2(x_3,\dots,z_1) \\
    &\vdots \\
    &z_n = x_n  \exp(s_n(z_1,\dots,z_{n-1})) + t_n(z_1,\dots,z_{n-1})
\end{split}
\end{align}}

\noindent where $s_i:\mathbb{R}^{n-1} \to \mathbb{R}$ and $t_i: \mathbb{R}^{n-1} \to \mathbb{R}$,  $i \in \{1,\dots,n\}$ are neural networks. Given $z = [z_1,\dots,z_n]$, $x$ is given by
{
\begin{align}
    x_n &= (z_n - t_n(z_1,\dots,z_{n-1})) \exp(-s_n(z_1,\dots,z_{n-1}))\nonumber\\
    x_{n-1} &= (z_{n-1} - t_{n-1}(x_n,\dots,z_{n-2}))\nonumber\\
    &\hspace{100pt} \exp(-s_{n-1}(x_n,\dots,z_{n-2}))\nonumber\\
    &\vdots \nonumber\\
    x_1 &= (z_1 - t_1(x_2,\dots,x_{n})) \exp(-s_1(x_2,\dots,x_{n})).\!
\end{align}}
The functions $s_i$ and $t_i$ can be arbitrarily chosen (e.g., not necessarily invertible). We implemented them to be neural networks comprised of fully connected layers with Sigmoid Linear Unit (SiLU) activations \cite{silu}. While one block of the above architecture is typically sufficient for representing $\phi_\theta$, multiple blocks can be stacked to increase the network's representation power.

To stabilize the training process, making the functions $s_i$ bounded (e.g., max absolute value = 1) is desirable as these are inputs to the $\exp$ function. To enable the reconstruction of larger values, we pass $z$ through another fully connected layer (i.e., $\bar{z} =   l(z) = W_lz + b_l$, which is a learnable scaling).
Since $W_l$ is a square matrix, constraints (such as loss function components) can be introduced to make it invertible and $z$ can be restored from $\bar{z}$ as $z = W_l^{-1}(\bar{z} - b_l)$.

\section{SIMULATION EXAMPLES}
\label{simul}
We apply our methodology to three nonlinear systems below and evaluate its efficacy. 
%% We demonstrate accuracy of the learned feedback linearization through comparisons between trajectories measured from the system (under random initial conditions and input signals) and trajectories estimated using the learned state transformations and linear dynamics. We also demonstrate that controller designs based on the learned linear system can be transformed back to the original system and that expected changes in closed-loop properties are indeed achieved. 
%and that expected modifications of the stability of the nonlinear system are indeed achieved, 
% therefore illustrating the validity of the learned feedback linearization. 
% \subsection{Training setup}

\subsection{Neural Network Training}
\label{subsec:inpsig}
To create training data (\ref{eqn:dataset}), we consider inputs of form $u(t) = \sum_{i=1}^{L}M_i\sin(\omega_it + \Phi_i) $,
% \begin{equation}
%     u(t) = \sum_{i=1}^{L}M_i\sin(\omega_it + \Phi_i) 
% \end{equation}
with $L=3$, $M_i \sim \mathcal{N}(1,1)$, $\omega_i \sim \mathcal{U}(0.5,1.5)$, and $\Phi_i \sim \mathcal{U}(-\pi,\pi))$ where $\mathcal{N}$ and $\mathcal{U}$ represent normal and uniform distributions respectively. After every $t_i$ seconds, the values of $M_i$, $\omega_i$, and $\Phi_i$ are re-sampled to increase the randomness of the input signal. To avoid discontinuities at $t_i$, we pass the signal through a low pass filter. We run simulations of the systems to create training data (\ref{eqn:dataset}) using several randomly sampled initial conditions. 
All the neural networks are trained offline with the collected trajectories using Pytorch and Adam optimizer with the initial learning rate set to $0.001$. 
%Training has been conducted on a PC with Intel Core i7 processor and 48 GB of RAM and Nvidia Titan XP GPU with 12GB of memory.   

% Using this input, we run simulations of the nonlinear systems at sampling frequency $f$ and sample several trajectories to create the dataset (\ref{eqn:dataset}). The subset $\mathbb{D}_q$ is defined as a single trajectory

\subsection{Evaluation Methodology}
\label{sec:eval}
In Section~\ref{results}, we apply our proposed algorithm for feedback linearization to the three example systems described in Section~\ref{sec:examples}. To evaluate the accuracy of the learned feedback linearizations in Section \ref{results}, we compare them against analytical solutions that assume knowledge of the system dynamics. 
In contrast to our method, the analytical solution is formed entirely in the continuous domain. Therefore, for simulations of analytical solutions, we use adaptive step size integration methods (e.g., adaptive Runge-Kutta integration) in which the linearizing controller, as well as the state transform, are calculated within a variable time step loop with the original state dynamics. If the original state trajectories were to be simulated beforehand to be later used in finding the controller and state transform (similar to Alg. \ref{alg:flus}), the analytical solution would be less accurate. 
Hence, in the comparisons below, we consider the best-case scenario for the analytical solution and show that the performance of the learned solution is comparable to the analytical solution (even though the learned solution is constrained to a fixed time step). Moreover, we show that in some cases, the learned solution performs better than the analytical solution (e.g., under parametric uncertainties since the learned solution inherently adapts to the uncertainties by learning from the actual system unlike the analytical solution).
% This is due to the fact that the fixed time step discretization in Algorithm \ref{alg:flus} creates a fixed time step integration process while calculating the values of the desired analytical functions, the accuracy of which depends on sampling frequency and integration errors. As our proposed method is based on learning the functions and matrices as described in previous sections, it learns to account for such integration errors and is more robust. 

Since our learning-based method does not require knowledge of the system dynamics, we study the effects of small uncertainties on the performance of the analytical solution for fairness of comparison. The magnitudes of these terms in the examples are chosen to be small to mimic typical parameter uncertainties that remain in real-world systems even when significant effort has been spent on system identification. With these small uncertainties, we study the resulting errors in the trajectories computed using the analytical feedback linearization and compare them against the errors with the learned  feedback linearization that does not require any {\em a priori} knowledge of the system dynamics.

\subsection{Example Systems}
\label{sec:examples}
\noindent{\em C1: Second order synthetic system:}
a nonlinear system 
\begin{align}
    \dot{x}_1 = x_2 + \delta_1 x_1^{3} + \delta_2 x_1 \ \ ; \ \ 
    &\dot{x}_2 = u + \delta_3 x_1 \sin(x_2)
\end{align}
where $\delta_1$, $\delta_2$, and $\delta_3$ are constant values. For our simulations, $\delta_1 = \delta_3 = -0.25$ and $\delta_2 = -0.01$. The training data consists of 30 trajectories of 2 seconds each at a sampling rate of 1 kHz, effectively giving us a total of 60,000 input-state pairs. For creating these trajectories, the initial conditions are uniformly picked for all states from  $(-10,10)$. The analytical solution for comparison is found using $\delta_1 = \delta_3 = -0.255$ and taking the term $\delta_2x_1$ to be unknown. 

\noindent{\em C2: Third order synthetic system:} a nonlinear system 
\begin{align}
    \dot{x}_1 &= \delta_1x_2 + \sin(x_1) \ \ ;\ \ 
    \dot{x}_2 = \delta_2x_3 + \delta_3 \sin(x_1)\cos(x_2) \nonumber\\
    \dot{x}_3 &= u -  x_1 x_2 \cos(x_3)
\end{align}
where $\delta_1$, $\delta_2$, and $\delta_3$ are constant values. For this example, we set $\delta_1 = \delta_2 = \delta_3 = 0.25$. The training data has 3500 trajectories of 2 seconds each at a sampling frequency of 200 Hz, thus giving us 1,400,000 input-states pairs. The initial conditions for trajectory generation are picked uniformly from the origin-centered cube of side 20. The analytical solution for comparison is found by considering $\delta_1 = \delta_2 = 0.27$ and assuming the term $\delta_3\sin(x_1)\cos(x_2)$ to be unknown. This change is not big enough to cause a significant difference in the general operation of the system as $\|\delta_3 \sin(x_1) \cos(x_1)\| \leq \|\delta_3\| \leq 0.25$ (for the original system), the value of which is not very high when compared to the general operating regions of the system.

\begin{figure}[h]
    \centerline{\includegraphics[scale = 0.15,trim=0in 0.8in 0in 1.4in, clip=true]{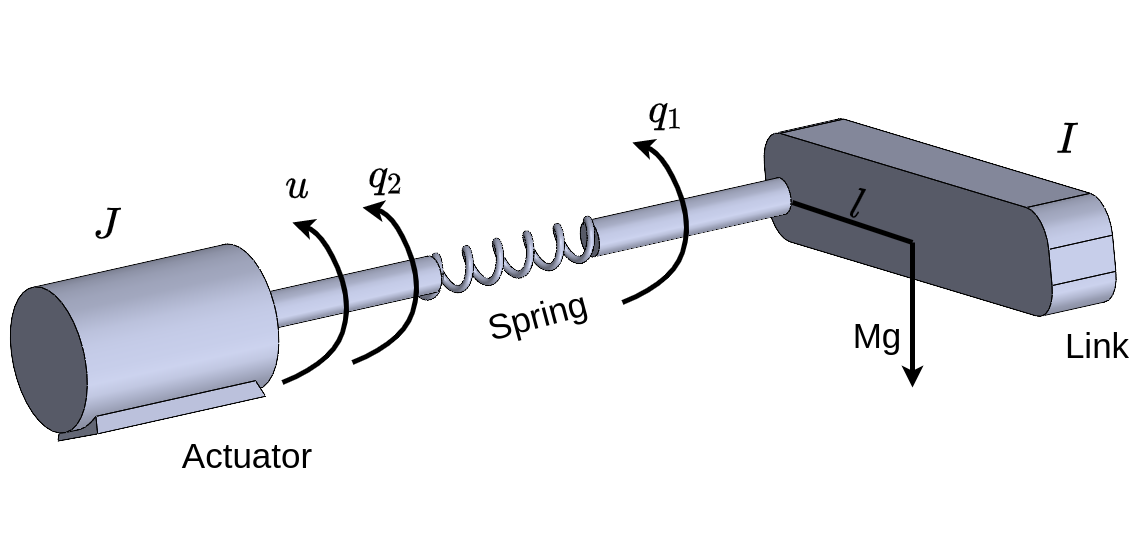}} 
    % \vspace*{-0.4cm}
    \caption{Single-link robot with flexible joint.}
    \label{fig:single-link}
\end{figure}

\noindent{\em C3: Single-link robot with flexible joint:}
% we also apply our algorithm to a simulation of a single-link robot with flexible joint (Fig. \ref{fig:single-link}) with joint friction and nonlinear spring. 
It is a fourth-order nonlinear system with a single input (Fig. \ref{fig:single-link}).
The dynamics with state variables given by angles $x_1=q_1$ and $x_3=q_2$ and angular velocities $x_2=\dot q_1$ and $x_4=\dot q_2$ are
\begin{align}
\begin{split}
    &\dot{x}_1 = x_2 \ \ ; \ \ 
    \dot{x}_2 = -\frac{Mgl}{I}\sin(x_1) - \frac{1}{I} \tau_{spring}\\
    &\dot{x}_3 = x_4\ \ ; \ \ 
    \dot{x}_4 = \frac{1}{J} \tau_{spring} + \frac{1}{J} u - \frac{1}{J}\tau_{friction}
\end{split}
\end{align}
where $M$ is the mass of the link, $g$ is gravitational acceleration, $l$ is the distance to the center of mass of the link from the point where the link is connected to the joint, $I$ and $J$ are moments of inertia of the link and actuator, respectively,  $\tau_{friction} = f_v {x}_4 + f_c sgn({x}_4) + (f_s - f_v)e^{-({x}_4/\omega_s)^2}$ is the joint frictional torque,  and $\tau_{spring} = \sum_{i=1}^3k_i sgn(x_1-x_3)^{i-1} (x_1-x_3)^i$ is the torque exerted by the  nonlinear spring. We use the following values: $M=1kg, l=1m, k_1 = 5Nm, k_2 = 2Nm, k_3 = 1Nm, g = 9.8m/s^2, I = 1kgm^2, J = 0.2kgm^2, f_v = 0.2Nms, f_c = 0.7Nm, f_s = 0.9Nm, \omega_s = 0.06 /s$ . We sample 1200 trajectories, each of 2 seconds, for the training data at a sampling frequency of 200Hz. For trajectory generation, initial conditions on $x_1$ and $x_3$ are picked uniformly from $(-\pi,\pi)$ and $x_2$ and $x_4$ from $(-3,3)$.  For training, we use larger weights (4x) for $x_1$ and $x_3$ compared to $x_2$ and $x_4$ to account for the typically smaller values of these states. The analytical solution is found by assuming that the frictional torque is zero.

\subsection{Results}
\label{results}
% For all the example systems considered above, it can be noted that the training loss converges to a minimum value in about 3000 epochs of training (Fig.~\ref{fig:epoch_data} (a)). 
% For improved convergence, we schedule the learning rate to decrease by a factor of 10 when the training loss reaches a plateau.
% Using the trained functions from Section \ref{subsec:learning}, we now evaluate the accuracy of the learned feedback linearization. 
% For this purpose, we consider an inference stage in which we measure the absolute errors in predicting the states in systems C1, C2, and C3.

\begin{figure}
    \begin{subfigure}[b]{.5\textwidth}
      \centering
          \includegraphics[trim=50 5 0 25,clip,scale = 0.215]{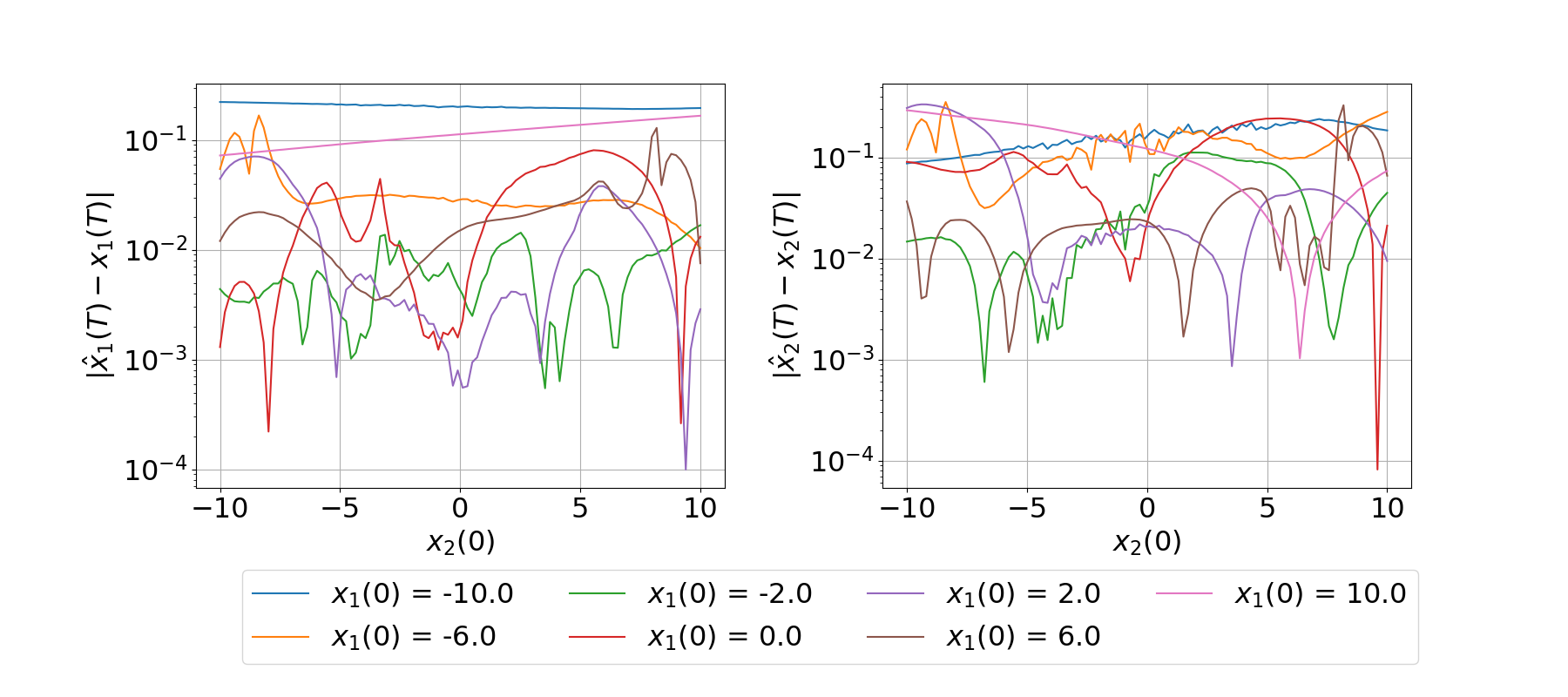}
          \caption{For each trajectory above, a constant value of the state $x_1$ is considered, while the value of the state $x_2$ has been varied.}
          \label{fig:sfig1}
    \end{subfigure}
    \begin{subfigure}[b]{.5\textwidth}
          \centering
          \includegraphics[trim=50 5 0 25,clip,scale = 0.215]{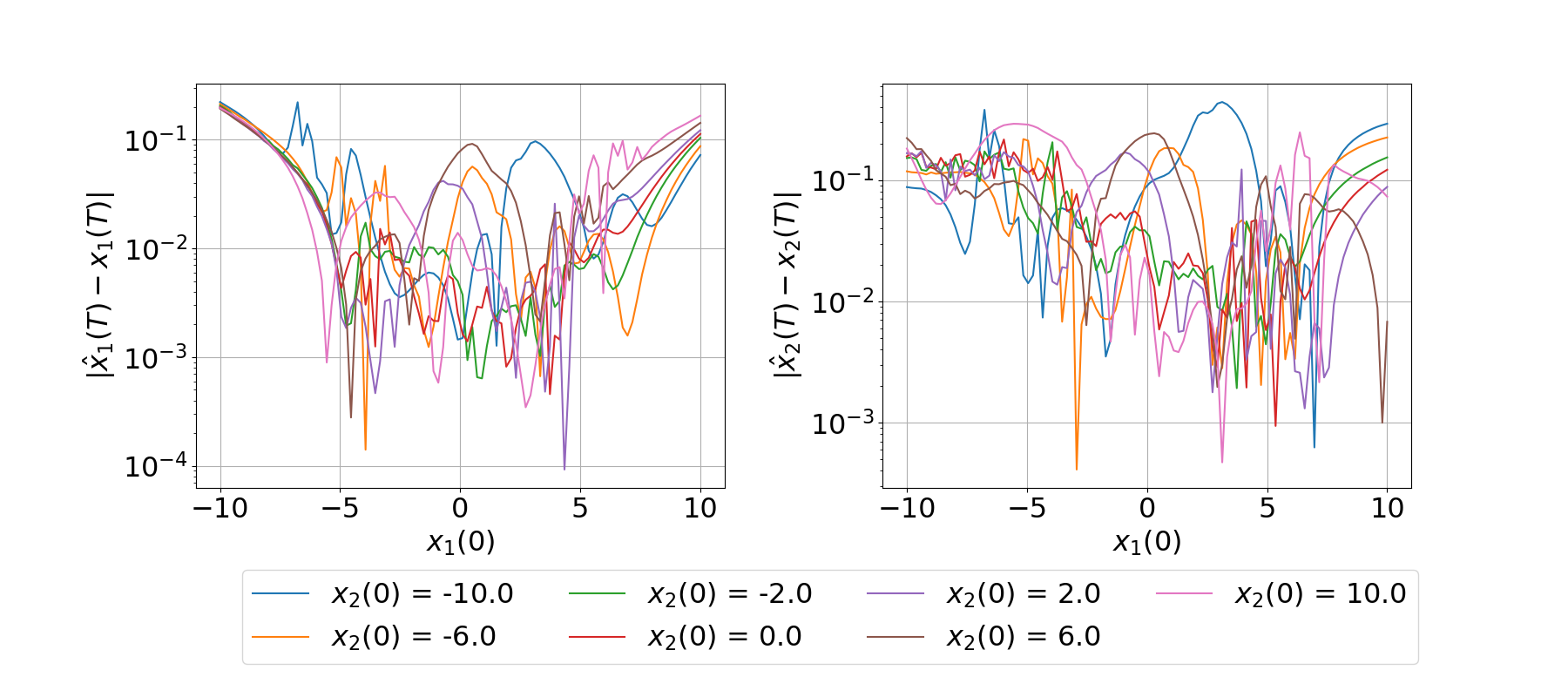}
          \caption{For each trajectory above, a constant value of the state $x_2$ is considered, while the value of the state $x_1$ has been varied.}
          \label{fig:sfig2}
    \end{subfigure}
    \caption{Single step ($T$ = 1ms) prediction error for system C1.}
    \label{fig:single_step}
\end{figure}

To evaluate the accuracy of the learned feedback linearization, we consider several initial conditions and input signals and measure the errors between the actual trajectories of the original systems and the {\em predicted} trajectories calculated using the learned feedback linearization. Since this effectively amounts to comparing solution trajectories of two different systems, we would expect that numerical inaccuracies and noise would generate a drift of increasing magnitude over time. This would be true of any solution approach and is observed also for analytical solutions with perfect knowledge of the system. Hence, what is to be compared between the learned and analytical solutions is the drift rate (i.e., the rate at which the errors grow). Before proceeding with this comparison, we first evaluate the ``instantaneous gradient'' behavior (i.e., single time step) of the learned solution to verify that the learned solution indeed reasonably matches the actual system.
% It is important to note that the errors in the plots in Figs.~\ref{fig:loss_eg1}, \ref{fig:loss_eg2}, and \ref{fig:loss_eg3} include the errors due to the open loop integration over multiple time steps which has been discussed in Section \ref{sec:eval}. Even with a very good approximation of the unknown functions, the errors are bound to increase over time due to the compounding of small integration errors. 
For this purpose, Fig.~\ref{fig:single_step} shows a solution snippet over a time horizon of 1 ms (i.e., a single time step in the discretized system representation) for the system in example C1. 
% As the sampling time used for this system is also 1 ms, this figure plots the prediction error over 1 time step, thus not allowing open loop integration errors to be added up. 
For this plot, we take different random values of the pair $(x_1(0), x_2(0))$ and $u(0)$ in our domain of interest and using our algorithm, find the values of $(\hat{x}_1(T),\hat{x}_2(T))$ which are then compared with the values of $(x_1(T),x_2(T))$ recorded from the system, with $T$ being 1 ms. 
% It is easy to observe that the average magnitudes of errors incurred in Fig.~\ref{fig:single_step} are significantly lower than that incurred in Figs.~\ref{fig:loss_eg1}, \ref{fig:loss_eg2}, and \ref{fig:loss_eg3}. 
It is seen that the errors are typically small indicating an ``instantaneous'' match between the actual system's trajectories and the predicted trajectories using the learned feedback linearization. 
Furthermore, as $(|x_1|,|x_2|)$ approaches $(10,10)$, the values of the errors in Fig.~\ref{fig:single_step} tend to be larger since the domain considered during training for this example is $[-10,10] \times [-10,10]$. 
% So, any state of the system going outside this domain will incur a larger error value.

\begin{figure}[!t]
\vspace*{0.1in}
    \centering
    \includegraphics[trim=50 5 0 25,clip,scale = 0.215]{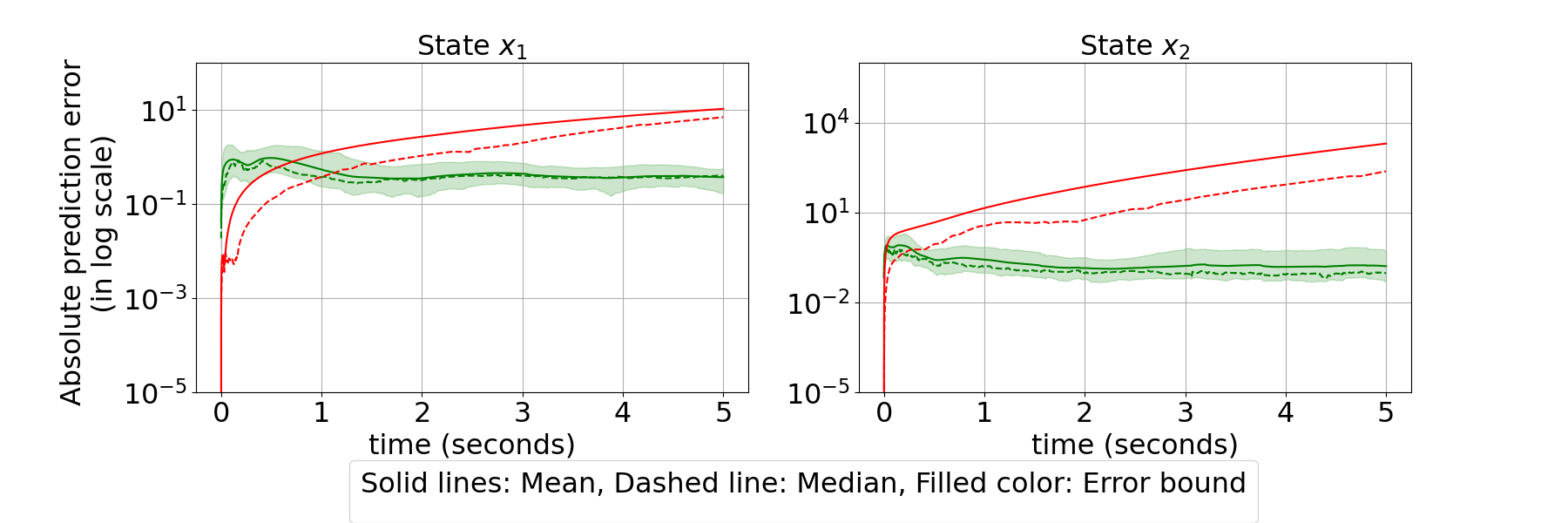}
    % \vspace*{-0.4cm}
    \caption{Errors between predicted/analytical and actual values of states of system C1. Green: learned solution; red: analytical solution (same notation is used for the subsequent plots).}
    \label{fig:loss_eg1}
\end{figure}

\begin{figure}[!t]
    \centering
    \includegraphics[trim=50 20 120 20,clip,scale = 0.21]{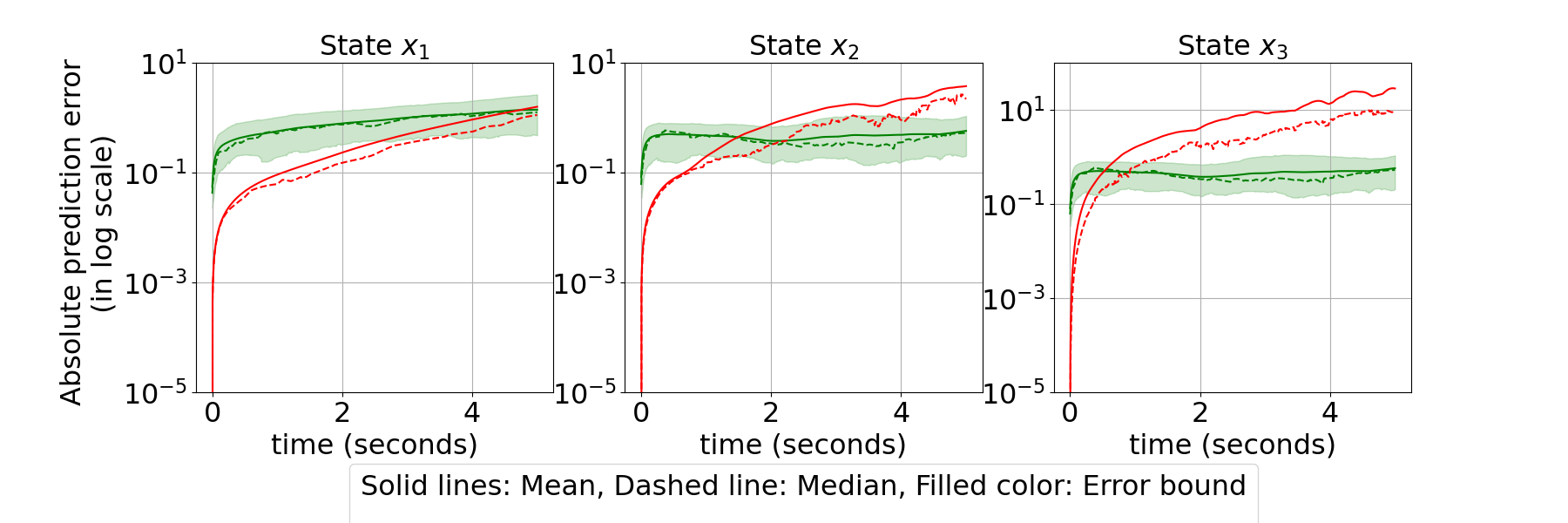}
    % \vspace*{-0.8cm}
    \caption{Errors between predicted/analytical and actual values of states of system C2.}
    \label{fig:loss_eg2}
\end{figure}

\begin{figure}[t]
\vspace*{0.1in}
    \centering
    \includegraphics[trim=45 10 120 45,clip,scale = 0.21]{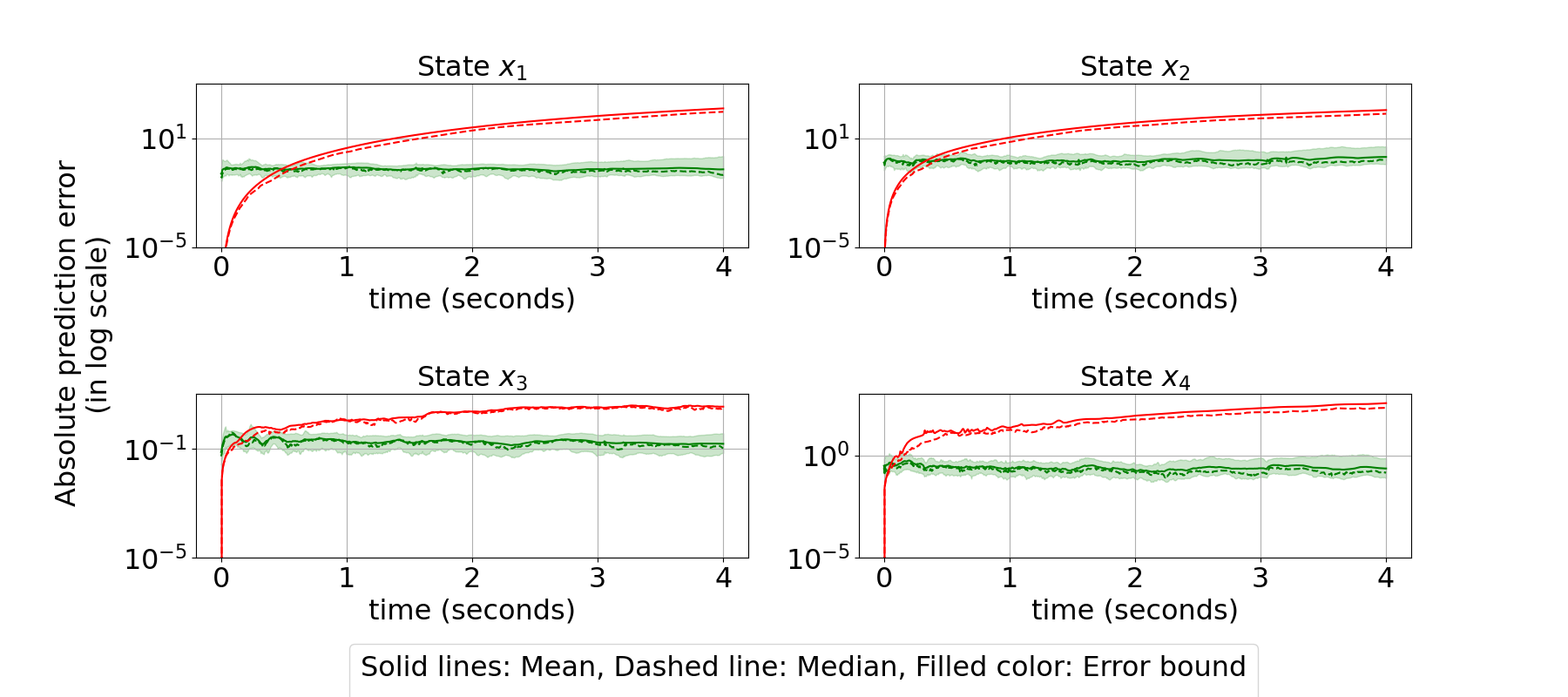}
    % \vspace*{-0.4cm}
    \caption{Errors between predicted/analytical and actual values of states of system C3. }
    \label{fig:loss_eg3}
\end{figure}

Now, we consider the errors between the actual and predicted trajectories over a longer time horizon. The absolute errors in predicting the states in systems C1, C2, and C3 are shown in Figs.~\ref{fig:loss_eg1}, \ref{fig:loss_eg2}, and \ref{fig:loss_eg3}, respectively. For each of the systems, 50 trajectories are sampled using inputs and initial conditions randomly generated as during the training.  Using the prediction part of Algorithm \ref{alg:flus}, we predict the values of the states and find the absolute errors. In Figs.~\ref{fig:loss_eg1}, \ref{fig:loss_eg2}, and \ref{fig:loss_eg3}, the error bound represents one standard deviation of the error below and above the mean value. 
To compare this performance with the analytical solution obtained with the uncertain dynamics described in Section \ref{sec:examples}, we evaluate the trajectories again but with $\alpha$, $\beta$, and $\phi$ calculated as described at the end of Section \ref{sec:i2sfl} with $f,g$ being the system dynamics (but with the uncertainties discussed in Section~\ref{sec:examples}), and $(A,B)$ being in the Brunovsky's form. The errors obtained with the analytical solution are also shown in Figs.~\ref{fig:loss_eg1}, \ref{fig:loss_eg2}, and \ref{fig:loss_eg3}. 
It is observed that the errors with the learned solution compare favorably with the analytical solution, showing the accuracy of the learned feedback linearization. Indeed, over longer time horizons, it is seen that even small parametric inaccuracies result in drifts with the analytical solution that are larger than the drifts with the learned solution (since the learned solution does not assume {\em a priori} knowledge of the dynamics and learns from state trajectories of the actual system).

% \begin{figure*}
%     \centering
%     \includegraphics[trim=70 50 100 50,clip,scale = 0.225]{loss.001.png}
%     \caption{Absolute errors between predicted and actual values of states of the example system C1 with respect to time.}
%     \label{fig:loss_eg1}
% \end{figure*}

% To evaluate the dependence of the performance on the amount of training data, we considered different sizes of the training dataset.
% Although increasing the size of the training dataset by sampling more trajectories from
% the nonlinear system generally improves the performance of the proposed algorithm, it was noted that after sampling a sufficient number of trajectories, sampling more data does not increase the performance significantly beyond a certain point (which conceptually corresponds to covering the addressed region of the state space sufficiently) as seen in Fig.~\ref{fig:epoch_data}.

% The learned $\alpha_\theta$ and $\beta_\theta$, $v$, and the predicted $x_\theta$ and $z$ for the single-link flexible-joint robot are given in Fig.~\ref{fig:eg} (as well as  the state trajectories $x$ from the original system). The input signal is generated as described in Section \ref{subsec:inpsig} and the initial conditions on the states are picked from the domain detailed in Section \ref{sec:examples}. It can be observed that the predicted state trajectories track the original trajectories with a small error, further indicating that the learned feedback linearizing functions are good approximations of the unknown actual functions.  In this example, we choose $\epsilon_3 = 100$ in the loss function in (\ref{l3}), because of which the range of $v$ has been restricted to within an absolute value of $100$. 

% \begin{figure}[!t]
% \vspace*{0.1in}
%     \centering
%     \includegraphics[trim=75 25 0 45,clip,scale = 0.220]{epoch_data_2.png}
%     % \vspace*{-0.4cm}
%     \caption{(a) Absolute prediction error against training epoch when 30 trajectories are considered in the training dataset for example system C1; (b) Absolute prediction error after convergence against the number of trajectories  for example system C1.}
%     \label{fig:epoch_data}
% \end{figure}

% \begin{figure}
%     \centering
%     \includegraphics[trim=95 55 120 125,clip,scale = 0.202]{9.png}
%     % \vspace*{-0.8cm}
%     \caption{Trajectories of learned functions $\alpha_\theta(x)$, $\beta_\theta(x)$, $x_\theta (= \phi^{-1}_\theta(z))$, states of linear system $z$, new input $v$, states of original system $x$,  and input $u$ for the system in C3. In this figure, $x_{i,\theta}$ represents the $i^{th}$ element of state $x_\theta$.}
%     \label{fig:eg}
% \end{figure}

% \begin{figure}[h]
% \vspace*{0.1in}
%     \centerline{\includegraphics[trim=0 5 80 20,clip,scale = 0.2]{loss_noisy_2.png}} 
%     % \vspace*{-0.2cm}
%     \caption{Effect of noise on performance.}
%     \label{fig:loss_noisy}
% \end{figure}

\begin{figure}[ht]
\vspace*{0.1in}
    \centerline{\includegraphics[trim=50 25 120 25,clip,scale = 0.2]{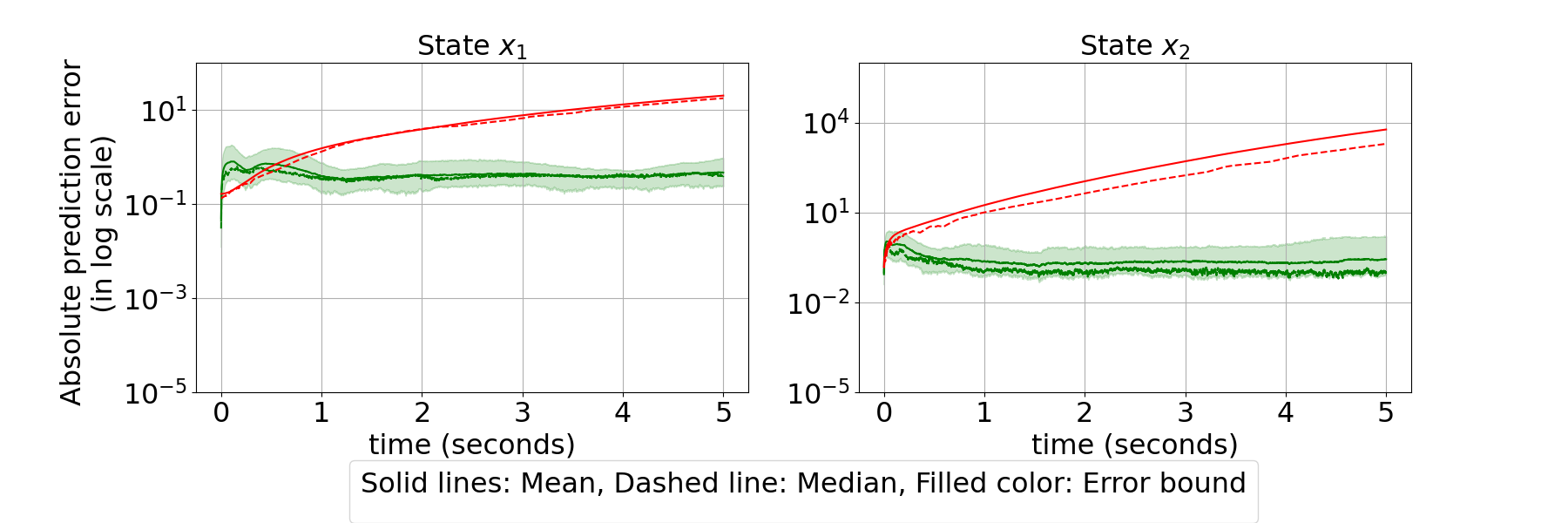}} 
    % \vspace*{-0.2cm}
    \caption{Effect of measurement noise on performance on C1. }
    \label{fig:mnoise}
\end{figure}

\begin{figure}[ht]
\vspace*{0.1in}
    \centerline{\includegraphics[trim=50 25 130 25,clip,scale = 0.2]{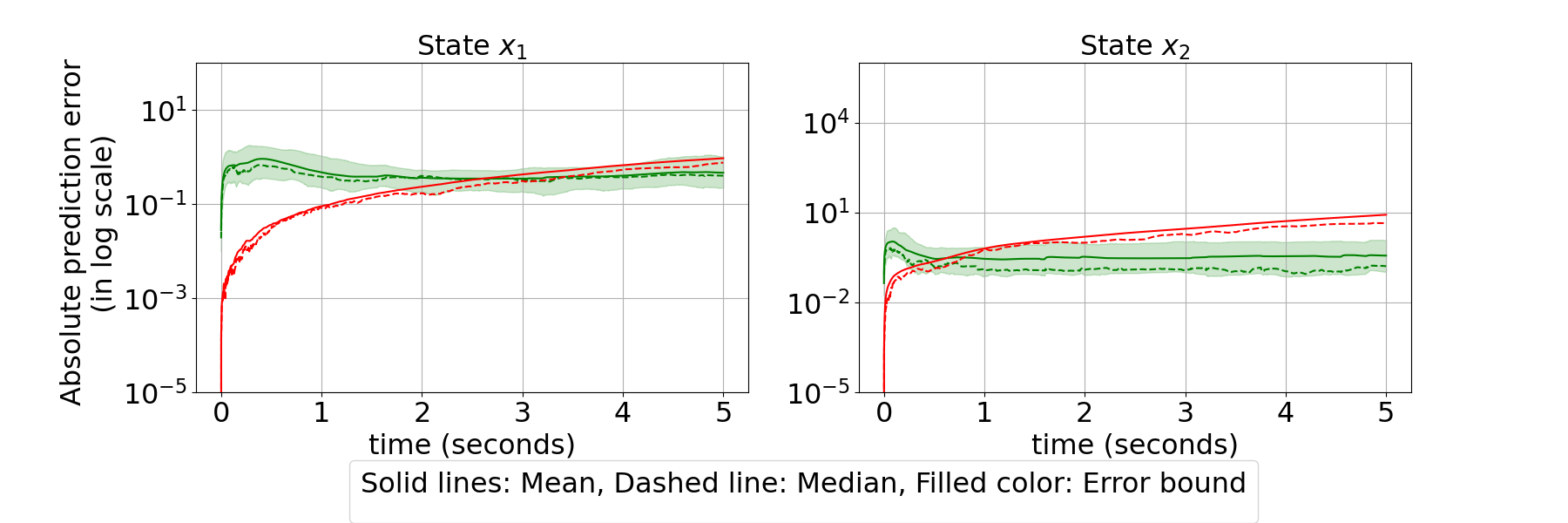}} 
    % \vspace*{-0.2cm}
    \caption{Effect of process noise on performance on C1.}
    \label{fig:pnoise}
\end{figure}

\subsection{Study of Robustness to Noise}
The simulations in Section \ref{results} assumed that the sensors used to measure states are perfect and that there is no process noise in the system model. We now consider the effects of measurement noise (Fig. \ref{fig:mnoise}) and process noise (Fig. \ref{fig:pnoise}). For these cases, Gaussian noise terms of variance 0.05 and mean 0 are added to the measurement and process models, respectively. The analytical solutions in these two cases assume full knowledge of the system dynamics apart from the random noise. The low magnitudes of the prediction errors as compared to the analytical solution shows that our proposed learning-based method actually provides better robustness to noise than the analytical solution. We hypothesize that the underlying reason for lower sensitivity to noise of the learning-based solution is that the learned $A_\theta$ matrix is typically a  stable matrix, resulting essentially in a low pass filtering effect that attenuates the random noise.
% In the simulation for system C1 in Fig.~\ref{fig:loss_noisy}, we add white Gaussian noise with variance 0.5 to the values of states and inputs during both training and inference stage. The low magnitude of prediction error shows the robustness of our proposed method.

\subsection{Stability Analysis}
\label{sec:stable}
\begin{figure}
    \centering
    \includegraphics[trim=120 40 120 100,clip=true, width = 0.8\linewidth]{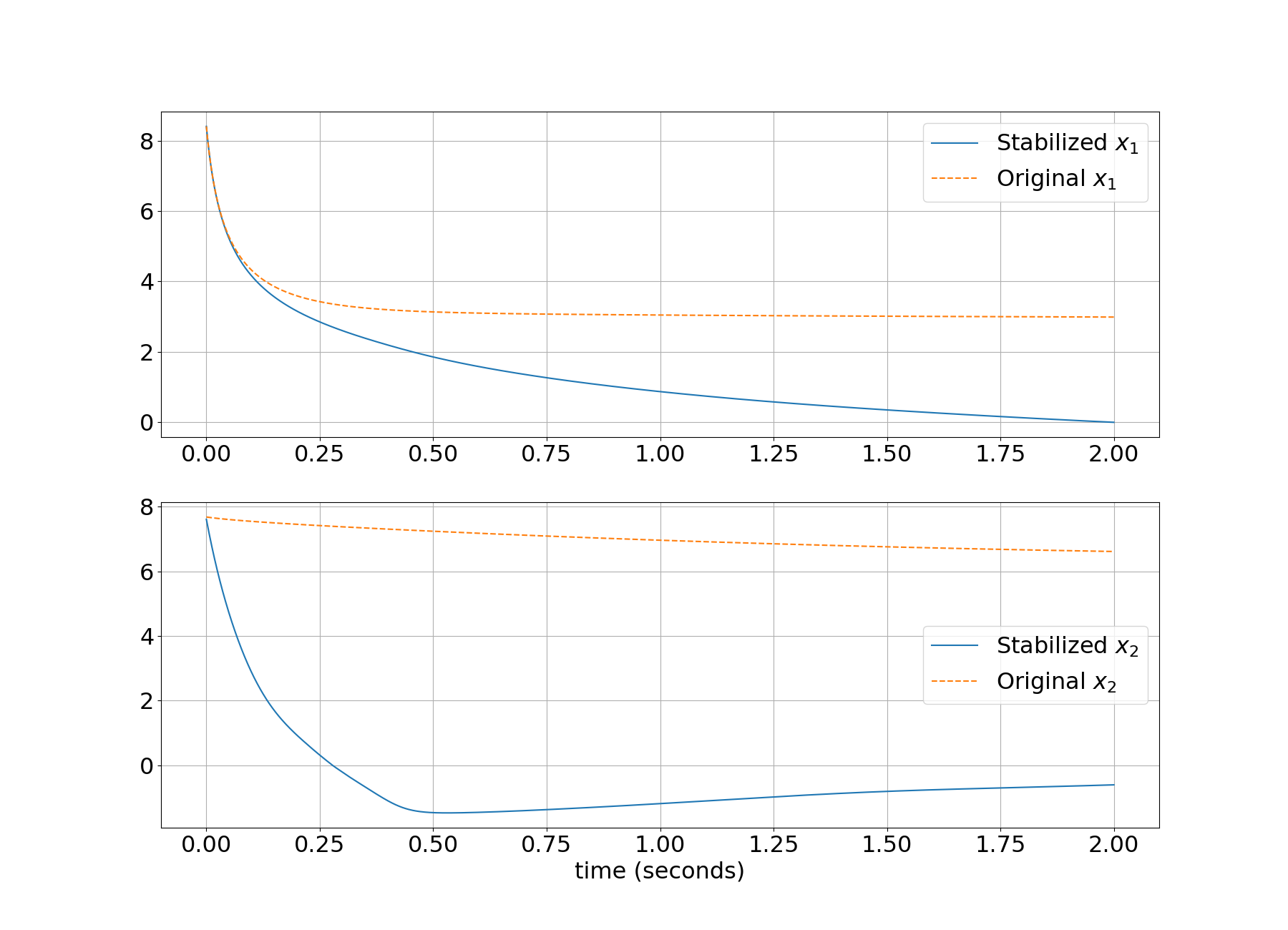}
    %\vspace*{-0.7cm}
     \caption{Demonstrating change in convergence rates of the systems with the  control law defined in Section \ref{sec:stable} for C1.}
    \label{fig:more_stable}
\end{figure}

Feedback linearization does not entail any constraint on the pair $(A,B)$ except controllability.
%and the overall system satisfies the constraints listed in Section \ref{subsec:model-uncert}. 
In our experiments, we have noted that the learned $A_\theta$ typically has all its poles inside the unit circle signifying that the network tends to learn a linear dynamic model that is  asymptotically stable.
%which implies that as $t \to \infty$, $z(t) \to 0$. 
But, this does not necessarily imply that $x_\theta(t) \to x_{eq}$ (an equilibrium point of the nonlinear system) since $\phi(x_{eq})$ is not necessarily 0. To this end, we augment the controller to achieve asymptotic convergence of the nonlinear system to $x_{eq}$.
Let $z_{x_{eq}} = \phi_\theta(x_{eq})$.
%where $x_{eq}$ is the desired equilibrium point of the nonlinear system. 
Therefore, we define the input to the linear system as the static state feedback $v = K(z-z_{x_{eq}})$ with $K$ picked to place the poles at the desired locations. 
%Noting $z_{x_{eq}}=\phi_\theta(x_{eq})$.
The modified controller for the nonlinear system is $u  = \alpha_\theta(x) + \beta_\theta(x)K(\phi_\theta(x)-\phi_\theta(x_{eq}))$.
% \begin{align}
%     %u &= \alpha_\theta(x) + \beta_\theta(x)K(z-z_{x_{eq}}) \\
%     u  &= \alpha_\theta(x) + \beta_\theta(x)K(\phi_\theta(x)-\phi_\theta(x_{eq})).
%     \label{eq:u_defn}
% \end{align}
Consider, for example, system C1, where, the desired equilibrium point is $x_{eq} = [0,0]^{T}$, which is already locally stable. Fig.~\ref{fig:more_stable} shows the system trajectories before (i.e., $v=0$) and after applying this control law (resulting in faster convergence).
%% It is seen the convergence faster using the control law defined above by suitably choosing $K$.
%% For the original system plots in Fig.~\ref{fig:more_stable}, we take $v(t)=0 \; \forall t$. We note that there is a decrease in the convergence time after the control input is applied. 
% Moreover, because we are applying $v = K(z-z_{x_{eq}})$ to the linear system, it now converges to $z_{x_{eq}}$ instead of 0, which is evident from the third and fourth plots in Fig.~\ref{fig:more_stable}. 

\begin{figure}[!t]
%\vspace*{0.1in}
    \centering
    \includegraphics[trim=70 5 120 25,clip,scale = 0.22]{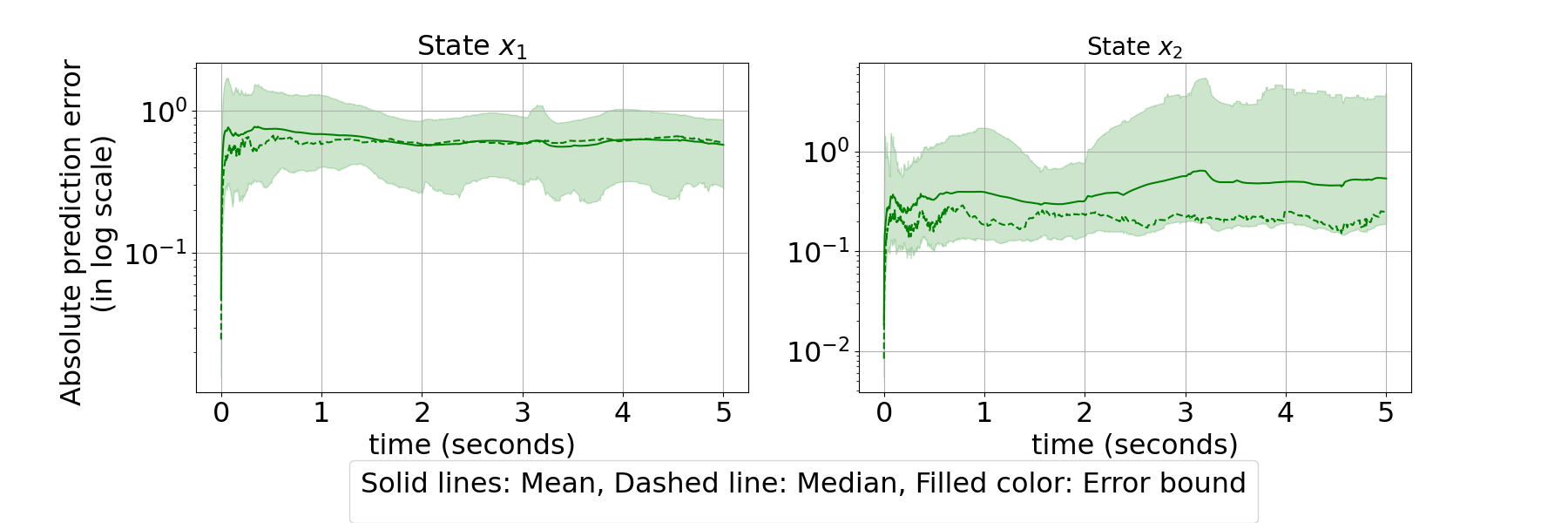}
    % \vspace*{-0.2cm}
    \caption{Absolute errors between predicted and actual states of C1 when trained with the loss function including $\mathcal{L}_4$.}
    \label{fig:eigen_loss}
\end{figure}

In our simulations, we find that sometimes the poles of the learned $A_\theta$ can be well inside the unit circle meaning that the learned feedback controller has a stabilizing effect on the transformed linear system. From a feedback linearization point of view, this is completely acceptable. However, if we wish to obtain an eigenvalue structure similar to Brunovsky's form in which all the discrete-time eigenvalues have magnitude 1, we can do so by adding another loss
$\mathcal{L}_4 = \big||\det{A_\theta}| - 1\big|$
to (\ref{loss}) with some weight $a_4$. This is based on the fact that the determinant of a matrix is the product of its eigenvalues. So, by using $\mathcal{L}_4$, we encourage $A_\theta$ to have eigenvalues as close to the unit circle as possible. While the learned linear system had poles at 0.9951 and 0.75 when not using $\mathcal{L}_4$, the inclusion of $\mathcal{L}_4$ results in poles at 0.9954 and 0.9881. Comparing Figs.~\ref{fig:eigen_loss} and \ref{fig:loss_eg1}, we note that there is no significant performance drop after using $\mathcal{L}_4$ although the poles are closer to the unit circle.

\section{CONCLUSIONS}
We proposed an algorithm that uses data-driven techniques to learn nonlinear state and input transformations to feedback linearize nonlinear systems with unknown dynamics. For this purpose, we developed novel techniques including a deep learning architecture to learn state and input transformations, loss functions capturing the feedback linearization objectives, and an extension of INNs for representing nonlinear changes of coordinates. 
% We also showed the application of the learned feedback linearization for designing controllers for the original nonlinear system. 
Future work will address application to multi-input systems, partial feedback linearization for systems that can not be completely feedback linearized,
and optimal sampling of initial conditions and input signals for generation of training data to further improve training convergence.
% In particular, we used deep learning based methods to estimate the linearizing controller and the nonlinear change of coordinates. Moreover, we proposed an extension of invertible neural network which helped us in calculating this nonlinear change of coordinates and defined loss functions to satisfy the desired constraints for feedback linearizing a system. Finally, we showed how a basic stabilizing controller can be constructed using the learned linearizing control functions. Future works in this domain can include extending this algorithm to a multiple input system and to systems giving zero order dynamics. Furthermore, a theoretical analysis on the stability of the system can be conducted using methods like Lyapunov stability theorems.


%%%%%%%%%%%%%%%%%%%%%%%%%%%%%%%%%%%%%%%%%%%%%%%%%%%%%%%%%%%%%%%%%%%%%%%%%%%%%%%%

\bibliographystyle{IEEEtran}

\bibliography{acc2023}

\end{document}